\documentclass[doublecol]{epl2}

\usepackage{graphicx}
\usepackage{amssymb,textcomp,txfonts}
\usepackage{soul}

\title{Synchronised firing induced by network dynamics in excitable systems}
\shorttitle{Synchronisation induced by network dynamics}

\author{Claudio J.~Tessone\inst{1} \and Dami\'an H. Zanette\inst{2} }
\shortauthor{C.J.~Tessone \& D.H.~Zanette}

\institute{
  \inst{1} Chair of Systems Design, D-MTEC, ETH Zurich. Kreuzplatz 5, CH-8032 Z\"urich, Switzerland\\
  \inst{2} Consejo Nacional de Investigaciones Cient\'{\i}ficas y
  T\'ecnicas, Centro At\'omico Bariloche, 8400 Bariloche, R\'{\i}o Negro, Argentina}


\abstract{We study the collective dynamics of an ensemble of coupled
identical FitzHugh--Nagumo elements in their excitable regime. We
show that collective firing, where all the elements perform their
individual firing cycle synchronously, can be induced by random
changes in the interaction pattern. Specifically, on a sparse
evolving network where, at any time, each element is connected with
at most one partner, collective firing occurs for intermediate
values of the rewiring frequency. Thus, network dynamics can replace
noise and connectivity in inducing this kind of self-organised
behaviour in highly disconnected systems which, otherwise, wouldn't
allow for the spreading of coherent evolution.}

\begin{document}

\maketitle

\section{Introduction}

The spontaneous emergence of coherent behaviour in populations of
interacting units ---be they of physical, chemical, biological, or
technological nature--- is crucial to their collective function.
Synchronisation of several kinds occurs in such disparate systems as
mechanical oscillators, bio-molecular reactions, neural networks,
insect societies, hormonal cycles, coupled lasers, and Josephson
junctions. Mathematical models for this variegated class of
phenomena have been proposed in terms of ensembles of coupled
dynamical systems of different types: linear and nonlinear
oscillators, chaotic elements, excitable units, among others
\cite{PRK:2001,MMZ:2004}.

It is a well-established fact that, in a population of interacting
elements, sufficiently strong, attractive coupling induces
self-organised synchronisation. This occurs even in the presence of
external noise, or when the individual behaviour of each element is
chaotic, or when elements are not identical to each other, with the
proviso that the population is well-interconnected in such a way
that information about the state of any element can reach any other.
While the structure of the interaction pattern can affect details in
the collective dynamics \cite{Wu:2007,ARENAS2008}, connectedness and
strong coupling generally guarantee synchronisation.

It has recently been shown that, both in synchronisation and in
contact processes (such as epidemics spreading), instantaneous lack
of connectivity can be compensated by dynamical rewiring of the
interaction network \cite{Vazquez2010a,Z:2011}. Specifically, in
populations with very sparse, disconnected instantaneous interaction
patterns, the respective transitions to full synchronisation and to
endemic states are triggered by increasing the rewiring rate. This
result is relevant, especially, to biological and social networks,
where potential contacts between the members of a population are not
continuously realised, but can occasionally be activated.

In this Letter, we disclose a related but different phenomenon,
concerning the collective dynamics of populations of excitable units
on evolving networks. Interacting excitable elements, which
individually perform a ``firing'' cycle in phase space if perturbed
strongly enough from their quiescent state, are known to undergo
collective synchronised firing induced by external noise
\cite{KS:1995,Tessone2007} and by repulsive interactions \cite{Tessone2008}.
We show here that, even in the absence of noise, an ensemble of
coupled FitzHugh--Nagumo excitable elements on an evolving, very
sparse, network exhibits collective firing for intermediate values
of the rewiring rate. This phenomenon in characterised numerically
for different network topologies and semi-quantitatively explained
in terms of the perturbations that network reconnections impose on
the individual dynamics of each element.

\section{Excitable elements on an evolving network} \label{sect:fhn}

FitzHugh--Nagumo excitable elements constitute an archetypical model
for type-II excitability, which occurs in many natural and
artificial systems ranging from epidemic spreading, to neural
and cardiac tissues \cite{CK:1999,GHmC:1991} to chemical reactions
and electronic devices \cite{LGNS:2004}.  The model is defined in
terms of an activatory (fast) variable  $x$ and an inhibitory (slow)
variable $y$. We consider an ensemble of $N$ sparsely connected
FitzHugh--Nagumo elements, whose dynamics is given by
\begin{eqnarray}
\label{eq:x} \epsilon\dot x_j& = & x_j-\frac{1}{3}x_j^3+y_j+ K_j(t)
\left(x_{j^*}-x_j \right),
\\
\label{eq:y} \dot y_j & = & a - x_j,
\end{eqnarray}
for $j=1,\dots,N$. The time-dependent factor $K_j(t)$ weights the
interaction between elements $j$ and $j^*$, as explained below. The
small parameter $\epsilon$ measures the time-scale ratio between the
fast and slow variables $x_j$ and $y_j$. The positive parameter $a$
characterises the dynamical regime of the non-interacting ($K \equiv
0$) element: for $a < 1$, it performs a periodic oscillation in the
$(x_j,y_j)$-plane, while for $a\geq 1$ its behaviour is excitable.
In this latter regime, and in the absence of external perturbations,
the non-interacting element asymptotically approaches the sole
stable fixed point $(x_{\rm eq},y_{\rm eq}) = (a, a^3/3 - a)$ and
remains quiescent there. Under a sizeable perturbation, however, the
element may exit the vicinity of the fixed point and return to it
after a long excursion in phase space ---usually referred to as a
firing cycle, or spike.

Our FitzHugh--Nagumo elements interact through a sparse evolving
network such that, at any given time, each element is coupled to
{\em at most} one partner. The coupling constant in eq.~(\ref{eq:x})
is $K_j(t)=k$ when element $j$ interacts with a generic partner
$j^*$ (not necessarily the same at all times), and $K_j(t)=0$ when
$j$ is isolated. This rather extreme sparseness determines, in a
sense, the most unfavourable situation for the emergence of
collective phenomena in an ensemble of interacting units. We expect
network dynamics to replace connectivity in triggering collective
behaviour.

In our numerical simulations, we study two different schemes for the
network dynamics. In the first one, the network consists of exactly
$N/2$ undirected links, distributed in such a way that every element
is always coupled to exactly one partner. As time elapses, two
connected pairs of elements are occasionally chosen at random to
mutually exchange their partners. Thus, two links in the network are
rewired.

The second scheme for network dynamics is built on top of an
underlying (undirected, connected) network $G$ with a fixed number
of links. During the dynamical evolution, however, only a subset of
the links is active. The links of the underlying network $G$ thus
represent the potential connections in the actual interaction
pattern. The initial network is generated by successively selecting
elements in a random order. If the chosen element is isolated, the
link to {\em one} of its still isolated neighbours in $G$ gets
activated. If no available neighbours exist, the element remains
with no active connection. During evolution, an inactive link from
$G$ is occasionally chosen at random and gets activated. At the same
time, pre-existing links of the newly connected elements become
deactivated.

It is not difficult to realise that, if there is no correlation
between the degrees of neighbour nodes in the underlying network
$G$, the frequency  $\omega_+ (z)$  with which an isolated node of
degree $z$ becomes connected, exactly equals the frequency $\omega_-
(z)$ with which it becomes isolated when it is connected. In turn,
the probability $P$ to find the node connected to any partner
satisfies
\begin{equation}
\dot P = \omega_+(z) (1-P)- \omega_-(z) P.
\end{equation}
Therefore, for asymptotically long times, $P=1/2$ for any $z$. In
other words, in our second reconnection scheme and for long times,
there are on the average $N/2$ connected elements ---and,
consequently, $N/4$ links--- at any time. The resulting interaction
pattern is thus twice as sparse as in the first scheme.

In both reconnection schemes, we denote by $\lambda$ the {\em
reconnection rate}, i.e.~the probability per time unit that the
partner of any given element changes.

\section{Order parameters}

To characterise the collective properties of our system in the
framework of the standard theory of synchronised oscillators
\cite{YK:1984}, it is convenient to compute a quantity describing
the {\em phase} of each excitable element along its firing cycle.
For the model defined by eqs.~(\ref{eq:x}) and (\ref{eq:y}), the
excursion in phase space occurs around the origin of the
$(x_j,y_j)$-plane. Thus, a suitable definition of the phase is
simply
 \begin{equation}
\phi_j(t) = \tan^{-1} \left[ \frac{y_j(t)}{x_j(t)} \right] .
\end{equation}

The behaviour of the ensemble, including possible transitions
between different collective dynamical regimes, can be statistically
characterised by a pair of order parameters defined in terms of the
individual phases $\phi_j(t)$ \cite{Tessone2007,Tessone2008}. First,
we take the average of the location of the particles on the unit
circle,
\begin{equation}\label{eq:rho}
\rho(t) \exp [ i \Psi(t)] = \frac{1}{N}\sum_{j=1}^N \exp [i \phi_j
(t)] ,
\end{equation}
and compute the Kuramoto order parameter as $\rho \equiv \langle
\rho(t) \rangle$, where $\langle \cdot \rangle$ stands for the time
average over a long time interval  \cite{YK:1984}. This parameter
measures the degree of synchronisation attained by the ensemble:
with full synchronisation we have $\rho=1$, whereas for a state
where phases are uniformly distributed over $[0,2\pi)$ we have $\rho
\sim N^{-1/2}$.

In excitable systems, however, the Kuramoto order parameter does not
allow to discern between the case where phases are statically
synchronised at the fixed point  $\phi_{\rm eq} = \tan^{-1 } (y_{\rm
eq} / x_{\rm eq})$, and the case where they rotate coherently, as
expected to occur in the regime of collective firing. To
discriminate between static and dynamic synchronisation, we compute
the Shinomoto--Kuramoto order parameter \cite{SK:1986},
\begin{equation}\label{eq:zeta}
\zeta = \left< \left| \rho(t) \exp [i\Psi(t)] - \left< \rho(t)\,
\exp [i\Psi(t)] \right> \right|  \right>,
\end{equation}
which differs from zero for synchronous firing only.

A third relevant order parameter, frequently used in the analysis of
stochastic transport \cite{PR:2002}, is the current, which we
compute as
\begin{equation}\label{eq:j}
J=\left< \left| \frac{1}{N}\sum_{j=1}^N \dot x_j(t) \right| \right>,
\end{equation}
i.e. as the time average of the absolute mean velocity along the
coordinate $x$. It gives a measure of the level of (not necessarily
synchronised) firing in the ensemble.

\section{Numerical results}

We have performed extensive computer simulations of the model
defined by eqs.~(\ref{eq:x}) and (\ref{eq:y}) for $a=1.02$
(excitable regime) and $k=1$, with the corresponding network
dynamics. The results presented here are qualitatively
representative of a broad parameter range in the same regime. Order
parameters were computed after the system reached a stationary
state, with no further changes in its dynamical behaviour.

\begin{figure}[t]
\begin{center}
\includegraphics[width=0.36\textwidth]{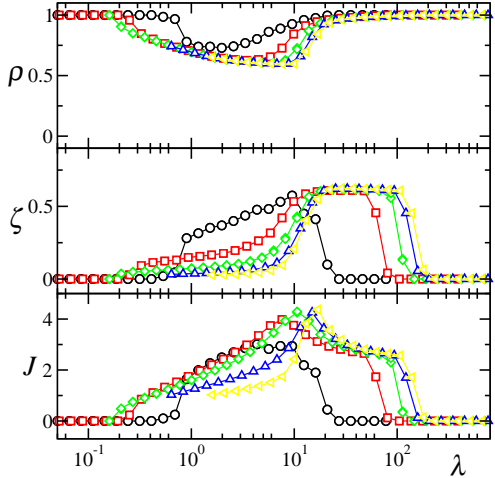}
\caption{ \label{fig:var_size} (Colour on-line) The order parameters
as functions of the reconnection rate $\lambda$, for the first
reconnection scheme, with  $k=1$, $a=1.02$, and $\epsilon=10^{-3}$.
Upper panel: the Kuramoto order parameter $\rho$; central panel: the
Shinomoto--Kuramoto order parameter $\zeta$; Lower panel: the
current $J$.  Different curves correspond to different system sizes:
$N=10$ $(\medcirc)$, $50$ $(\square)$, $200$ $(\diamondsuit)$, $800$
$(\triangle)$, and $2000$ $(\lhd)$. Joining lines are plotted as a
guide to the eye.}
\end{center}
\end{figure}

In fig.~\ref{fig:var_size}, we show numerical results for the order
parameters as functions of the reconnection rate $\lambda$, for the
first reconnection scheme and various system sizes. Both for small
and large values of $\lambda$, the Kuramoto order parameter is equal
to one, while the Shinomoto--Kuramoto order parameter and the
current vanish. This situation corresponds to a state where all the
elements of the ensemble are at rest at the same point in phase
space, namely, the fixed point $(x_{\rm eq},y_{\rm eq})$.

For intermediate reconnection rates, on the other hand, we find an
interval where the Kuramoto order parameter $\rho < 1$, indicating
that the elements are distributed over phase space. Within the same
interval, both the Shinomoto--Kuramoto order parameter $\zeta$ and
the current $J$ become positive and attain considerably high maxima.
This is an indication of collective firing with a concurrent
phase-space flow, and constitutes our main finding: reconnection
events at intermediate rates induce self-organised coherent
behaviour in an otherwise disconnected ensemble of FitzHugh--Nagumo
excitable elements.

Figure \ref{fig:var_size} also shows that the order parameters
become independent of the system size as $N$ grows. This suggests
that the regime of collective firing exists even in the
thermodynamic limit.

\begin{figure}[t]
\begin{center}
\includegraphics[width=0.36\textwidth]{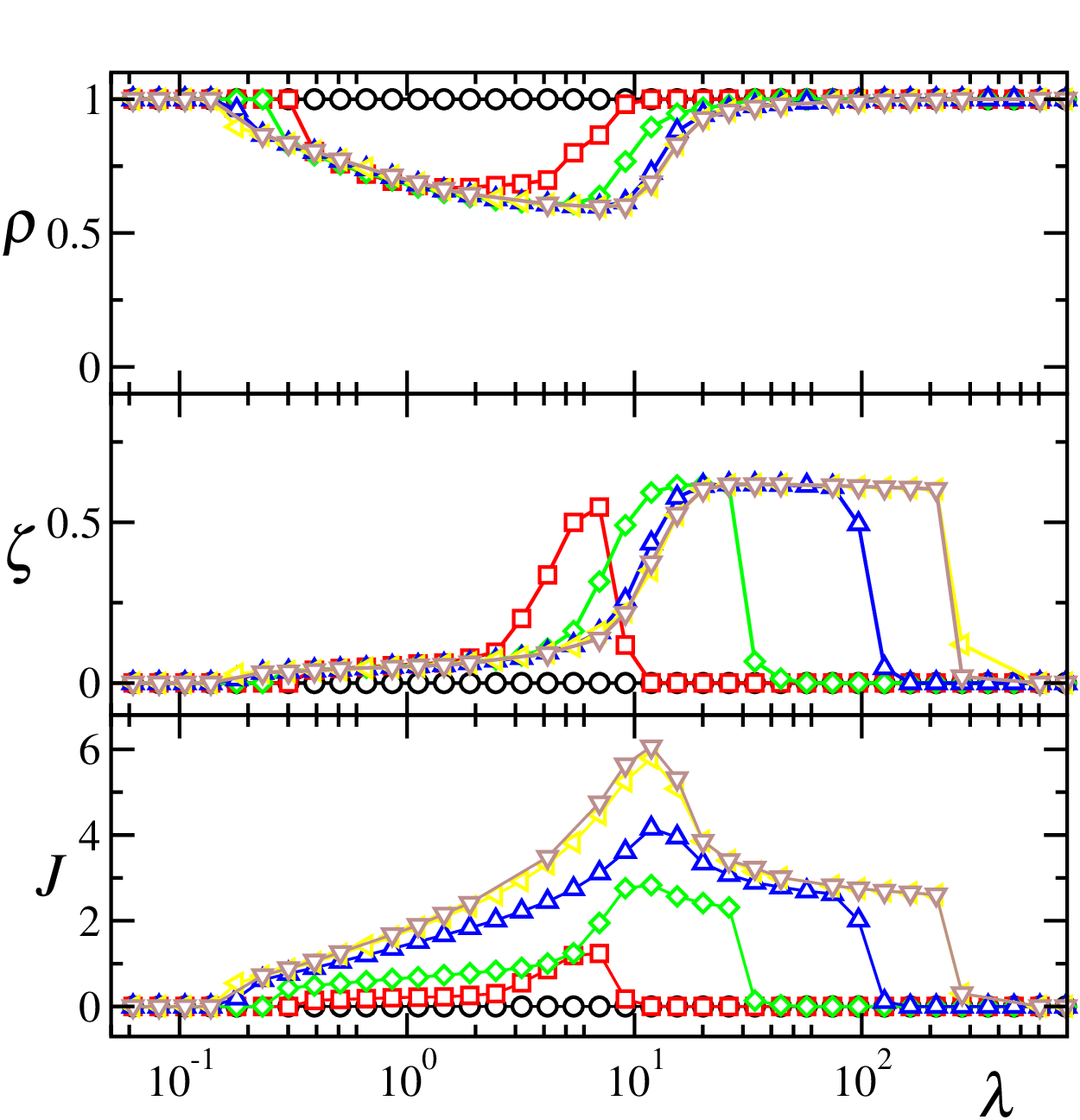}
\caption{ \label{fig:var_epsilon} (Colour on-line) As in
fig.~\ref{fig:var_size}, for a system of size $N=400$ and different
values of the time-scale ratio: $\epsilon=1$ $(\medcirc)$,
$10^{-1}$ $(\square)$, $10^{-2}$ $(\diamondsuit)$, $10^{-3}$
$(\triangle)$, $10^{-4}$ $(\lhd)$, and $10^{-5}$
$(\bigtriangledown)$.}
\end{center}
\end{figure}

Figure \ref{fig:var_epsilon} displays the order parameters for
different values of the time-scale ratio $\epsilon$ in a system of
size $N=400$. These results show that, when there is no difference
in the time scales associated to the variables $x$ and $y$
($\epsilon = 1$), collective firing is absent and all the elements
remain quiescent at the fixed point. As $\epsilon$ decreases,
however, the phenomenon takes place for intermediate values of
$\lambda$ and seems to approach a well-defined limit for $\epsilon
\to 0$. We provide a semi-quantitative analysis of this limit in the
next section.

To relax the condition that every element has a partner at any time,
we have used our second reconnection scheme with two kinds of
topologies for the underlying network $G$. We recall that, with this
scheme, the resulting instantaneous interaction pattern is more
sparse than in the previous case. Firstly, we have considered a
scale-free underlying network generated by the preferential
attachment rule \cite{BA:1999}, where each added node is connected
to $m$ pre-existing nodes. Secondly, we have taken a small-world
network built up from the rewiring, with probability $p$, of the
links of a two-dimensional network with Moore neighbourhood,
following the Watts--Strogatz prescription \cite{WS:1998}. We have
numerically verified that, as advanced above, the average number of
connected elements at long times fluctuates around $N/2$. For the
scale-free networks, this number is slightly, but systematically,
larger, which can be attributed to spurious degree--degree
correlations in the highly heterogeneous degree distribution
generated by preferential attachment in our finite-size system.

\begin{figure}[t]
\begin{center}
\includegraphics[width=0.36\textwidth]{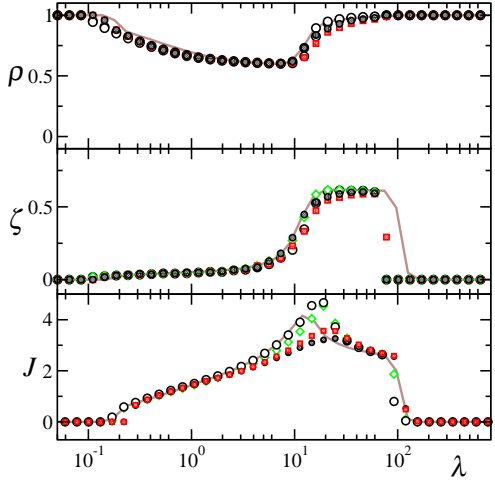}
\caption{\label{fig:var_topol} (Colour on-line) As in
fig.~\ref{fig:var_size} for the second reconnection scheme, with
$\epsilon =10^{-3}$ and $N=400$. Open symbols: scale-free underlying
network generated by preferential attachment with $m=3$
$(\diamondsuit)$ and $10$ $(\square)$. Full symbols: small-world
underlying networks generated by rewiring with  $p=0.01$
$(\medbullet)$ and  $0.1$ $(\blacksquare)$. The solid line
corresponds to the results for the first reconnection scheme.}
\end{center}
\end{figure}

Figure \ref{fig:var_topol} shows the order parameters for a system
of size $N=400$, underlain by scale-free networks with two values of
$m$ and small-world networks with two values of $p$. Solid curves
stand for the corresponding results for the first reconnection
scheme. Overall, the results are largely independent of both the
reconnection scheme and the topology of the underlying network and,
consequently, of the number of connected elements.

\section{Interpretation}

Whereas a full analytical description of synchronised firing in
dynamical networks of FitzHugh--Nagumo excitable elements seems to
be out of reach, it is possible to sketch a semi-quantitative
picture that plausibly explains the occurrence of this collective
phenomenon for intermediate values of the network reconnection rate.
The following arguments focus on our first reconnection scheme, but
can be straightforwardly extended to the second.

Consider eqs.~(\ref{eq:x}) and (\ref{eq:y}) for $\epsilon \to 0$. In
this limit, and in the absence of interactions ($k=0$), the fast
variable $x_j (t)$ follows adiabatically the slow variable $y_j (t)$
along the nullcline $\dot x_j =0$. Let us introduce the auxiliary
variable
\begin{equation} \label{eq:eta}
\eta_j (x_j) = \frac{1}{3} x_j^3 -x_j
\end{equation}
which, along the nullcline, satisfies $\eta_j = y_j + k
\left(x_{j^*}-x_j \right)$. For $k=0$, we have $y_j \equiv \eta_j$.
Differentiating with respect to time and taking into account
eq.~(\ref{eq:y}) yields
\begin{equation} \label{eq:etap}
\dot \eta_j = a - x_j(\eta_j) +k \xi_j,
\end{equation}
with $\xi= \dot x_{j^*}-\dot x_j$ and $x_j(\eta_j)$ given by the
inverse of the function in eq.~(\ref{eq:eta}). Note that
$x_j(\eta_j)$ is defined piecewise, depending on how $x_j$ compares
with $\pm 1$.

\begin{figure}[t]
\begin{center}
\includegraphics[width=0.36\textwidth]{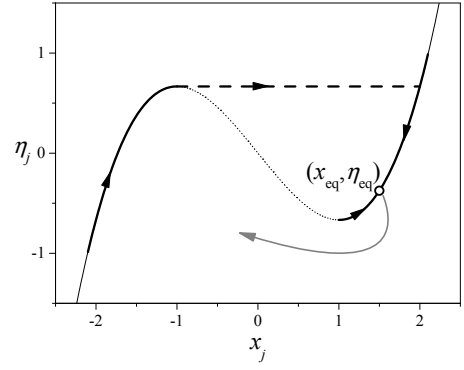}
\caption{\label{fig:ana}Phase-space dynamics of a single non-interacting
FitzHugh--Nagumo excitable element $j$ on the $(x_j,\eta_j)$ plane.
The narrow line represents eq.~(\ref{eq:eta}). Arrowed bold lines
are possible trajectories of the element. The fixed point $(x_{\rm
eq}, \eta_{\rm eq})$ is represented as an empty dot. The grey arrow
schematises the effect of a perturbation from the fixed point toward
negative values of $\eta_j$. }
\end{center}

\end{figure}

The arrowed bold lines in fig.~\ref{fig:ana} represent the
phase-space trajectories of an non-interacting element. In the limit
$\epsilon \to 0$, it is always found on the stable branches (either
$x_j<-1$ or $x_j>1$) of the nullcline, and asymptotically approaches
the fixed point at $(x_{\rm eq}, \eta_{\rm eq}) =(a , a^3/3-a)$,
plotted in the figure as an empty dot. If, as illustrated by the
grey arrow, the element is perturbed from the fixed point toward
negative values of $\eta_j$ and beyond the minimum $\eta_j (x_j=1) =
-2/3$, it immediately reaches the leftmost stable branch and begins
its excursion upwards. When it reaches $\eta_j (x_j=-1) = 2/3$, it
jumps to the rightmost branch and, from then on, it moves toward the
fixed point. The firing cycle is thus completed. Integration of
eq.~(\ref{eq:etap}) with $k=0$ and $a \gtrsim 1$ shows that, if the
leftmost branch is reached at $\eta_j \approx -2/3$, the time spent
on that branch is $\tau_{\rm left} = 1/2 + {\cal O}[a-1]$. In turn,
the typical time for relaxation toward the fixed point on the
rightmost branch is $\tau_{\rm right}= a^2 -1$.

Consider now the effect of interaction ($k \neq 0$) on the
individual dynamics of element $j$. If the reconnection rate
$\lambda$ is sufficiently small, so that $\lambda^{-1} \gg \tau_{\rm
left}, \tau_{\rm right}$, element $j$  remains connected to the same
partner $j^*$ over times which are long as compared with the typical
time scales needed to reach the vicinity of $(x_{\rm eq}, \eta_{\rm
eq})$. Irrespectively of the value of $k$, the two coupled elements
approach the fixed point well before their mutual link breaks and
they are reconnected to different partners. When reconnection
finally happens, however, all elements will be found near the fixed
point and the change of partner will have essentially no effect on
the subsequent dynamics of $j$. Therefore, the whole ensemble
converges to $(x_{\rm eq}, \eta_{\rm eq})$ over times of order
$\tau_{\rm left}+ \tau_{\rm right}$ and remains  there indefinitely.
For sufficiently small $\lambda$, hence, sustained collective firing
is absent.

As $\lambda$ grows and reconnection becomes more frequent, the term
$k \xi_j$ in the right-hand side of eq.~(\ref{eq:etap}) acquires the
character of a fluctuating force, analogous to additive noise. Since
$\xi_j (t) = \dot x_{j^*}(t) - \dot x_j (t)$, its time dependence
consist of a relatively smooth variation along the periods where
element $j$'s partner $j^*$ does not change, punctuated by sharp
delta-like ``kicks'' when reconnection occurs. Even if $j$ has
already reached the vicinity of $(x_{\rm eq}, \eta_{\rm eq})$, a
kick due to reconnection with an element which is transiting the
leftmost branch region may force $j$ to move away from the fixed
point and reinitiate its firing cycle. This event is schematised by
the grey arrow in fig.~\ref{fig:ana}. At appropriate values of the
reconnection rate, with most of the ensemble near $(x_{\rm eq},
\eta_{\rm eq})$, just a few ``outliers'' along the firing cycle are
able to induce a cascade of transitions from the fixed point to the
cycle, and collective firing is thus triggered. Our numerical
results show that, precisely, collective firing occurs for $\lambda
\gtrsim 1 \sim  (\tau_{\rm left}+ \tau_{\rm right})^{-1}$.

If reconnection grows even more frequent, within the time scales
relevant to the dynamics of a single element, the ``noise'' term $k
\xi_j$ averages out to its mean value over the whose ensemble.
Therefore, each element is effectively subject to the action of the
average state of the ensemble. In this situation, the interaction
between elements is equivalent to global (all-to-all) coupling
\cite{Vazquez2010a,Z:2011}. Since all the elements are identical,
global coupling leads the ensemble to collapse to the fixed point
\cite{MMZ:2004}, and collective firing is thus suppressed.

\begin{figure}[t]
\begin{center}
\includegraphics[width=0.42\textwidth]{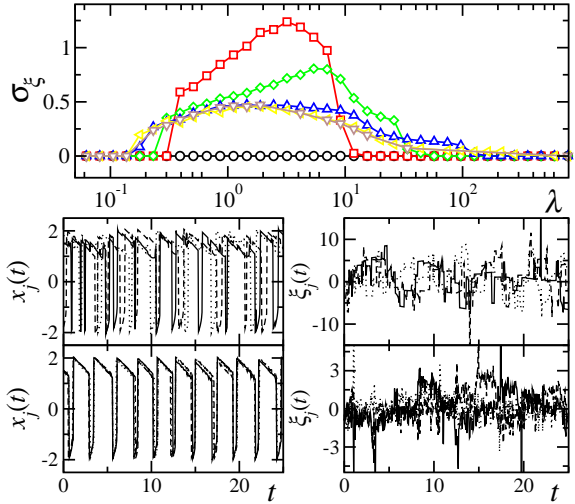}
\caption{\label{fig:effective_noise} (Colour on-line) Upper panel:
Standard deviation $\sigma_\xi$ of the ``noise'' $\xi_j (t)$ as a
function of the reconnection rate $\lambda$, for $k=1$, $a=1.02$,
$N=400$, and different values of the of the time-scale ratio:
$\epsilon=1$ $(\medcirc)$, $10^{-1}$ $(\square)$, $10^{-2}$
$(\diamondsuit)$, $10^{-3}$ $(\triangle)$, $10^{-4}$ $(\lhd)$, and
$10^{-5}$ $(\bigtriangledown)$. Lower panels: Time evolution of the
coordinate $x_j$ and the ``noise'' $\xi_j$ of three randomly
selected elements, for $\epsilon = 10^{-3}$, with $\lambda=0.32$
(upper row) and $\lambda=3.2$ (lower row).}
\end{center}
\end{figure}

The upper panel of fig.~\ref{fig:effective_noise} shows the standard
deviation $\sigma_\xi$ of the ``noise'' $\xi_j (t)$, averaged over
the ensemble and over time, as a function of the reconnection rate
$\lambda$ and for various values of the time-scale ratio $\epsilon$.
The lower panels show the coordinate $x_j(t)$ and the ``noise''
$\xi_j (t)$ as functions of time for a few selected elements, and
two values of the reconnection rate: $\lambda=0.32$ (upper row),
which corresponds to the threshold of global firing, and
$\lambda=3.2$ (lower row), where global firing is well developed.
For the latter, the synchronous pulsing of the coordinate $x_j(t)$
is apparent.

\section{Conclusion}

Synchronised collective firing in ensembles of coupled excitable
elements was known to be triggered by external noise
\cite{KS:1995,Tessone2007} and by disorder in the interaction
pattern \cite{Tessone2008} ---in this latter case, due to the
simultaneous presence of attractive and repulsive interactions. In
both situations, the emergence of this form of collective behaviour
requires the intensity of noise or the degree of disorder to be
neither too small nor too high: it is at an intermediate level of
fluctuations that the system has the appropriate dynamical
flexibility as to self-organise into coherent evolution.

In this Letter, we have shown that, in the absence of external
noise, the fluctuations associated with network dynamics ---when the
interaction pattern is rewired with a certain frequency--- are as
well able to induce collective firing of coupled excitable elements.
As in the previous instances, coherent evolution is observed for
intermediate values of the rewiring frequency. In the present
situation, network dynamics has the crucial additional role of
replacing the connectivity necessary to warrant the spreading of
information about the individual states of the excitable elements
all over the ensemble. In fact, by construction, the instantaneous
interaction pattern is highly diluted, with one or less neighbour
connected to each element at any time. This effect of network
dynamics had already been pointed out in chaotic synchronisation and
in contact processes \cite{Vazquez2010a,Z:2011}. Our results suggest
that the phenomenon of collective firing is remarkably independent
of the underlying structure of interactions and has a well-defined
behaviour in the thermodynamic limit of infinitely large
ensembles.

The present analysis pertains to the study of the ample variety of
systems where interaction patterns are not static, but change with
time either driven by external influences or in response to the
state of the system itself, or as a combination of both effects.
While this important dynamical aspect of complex systems has often
been disregarded, our results ---among other recent work---
highlight its role in the emergence of self-organised collective
evolution.

\acknowledgments   CJT acknowledges
financial support from Swiss National Science Foundation through grant CR12I1\_125298 and SBF (Swiss Confederation) through research project
C09.0055.  

\bibliographystyle{eplbib}
\bibliography{refs}

\begin{thebibliography}{10}
\expandafter\ifx\csname url\endcsname\relax\def\url#1{\texttt{#1}}\fi

\bibitem{PRK:2001}
\Name{Pikovsky A., Rosenblum M. \and Kurths J.} \Book{Synchronization: A
  universal concept in nonlinear sciences} 1st Edition (Cambridge University
  Press) 2001.

\bibitem{MMZ:2004}
\Name{Manrubia S., Mikhailov A. \and Zanette D.} \Book{Emergence of Dynamical
  Order: Synchronization Phenomena in Complex Systems} 1st Edition (World
  Scientific, Singapore) 2004.

\bibitem{Wu:2007}
\Name{Wu C.~W.} \Book{Synchronization in Complex Networks of Nonlinear
  Dynamical Systems} 1st Edition (World Scientific, Singapore) 2007.

\bibitem{ARENAS2008}
\Name{Arenas A., D\'{\i}az-Guilera A., Kurths J., Moreno Y. \and Zhou C.}
  \REVIEW{Physics Reports}{469}{2008}{93}.

\bibitem{Vazquez2010a}
\Name{Vazquez F. \and Zanette D.~H.} \REVIEW{Physica D: Nonlinear
  Phenomena}{239}{2010}{1922}.

\bibitem{Z:2011}
\Name{Zanette D.} \REVIEW{Pap. Phys.}{3}{2011}{030001}.

\bibitem{KS:1995}
\Name{{C. Kurrer} \and {K. Schulten}} \REVIEW{Phys. Rev. E}{51}{1995}{6213}.

\bibitem{Tessone2007}
\Name{Tessone C.~J., Scir\`{e} A., Toral R. \and Colet P.} \REVIEW{Physical
  Review E}{75}{2007}{1}.

\bibitem{Tessone2008}
\Name{Tessone C.~J., Zanette D.~H. \and Toral R.} \REVIEW{The European Physical
  Journal B}{62}{2008}{319}.

\bibitem{CK:1999}
\Name{Koch C.} \Book{Biophysics of Computation} (Oxford University Press, New
  York) 1999.

\bibitem{GHmC:1991}
\Name{Glass L., Hunter P. \and McCulloch A.} \Book{Theory of Heart}
  (Springer--Verlag, Berlin) 1991.

\bibitem{LGNS:2004}
\Name{Lindner B., Garc\'{\i}a-Ojalvo J., Neiman A. \and Schimansky-Geier L.}
  \REVIEW{Physics Reports}{392}{2004}{321}.

\bibitem{YK:1984}
\Name{Kuramoto Y.} \Book{Chemical Oscillations, Waves, and Turbulence}
  (Springer--Verlag, New York) 1984.

\bibitem{SK:1986}
\Name{Shinomoto S. \and Kuramoto Y.} \REVIEW{Prog. Theor.
  Phys.}{75}{1986}{1105}.

\bibitem{PR:2002}
\Name{Reimann P.} \REVIEW{Phys. Rep.}{361}{2002}{57}.

\bibitem{BA:1999}
\Name{Barab\'asi A. \and Albert R.} \REVIEW{Science}{286}{1999}{509}.

\bibitem{WS:1998}
\Name{Watts D. \and Strogatz S.} \REVIEW{Nature}{393}{1998}{440}.

\end{thebibliography}

\end{document}